\documentstyle[12pt]{article}
\def\lapprox{\mathrel{\smash{\mathop{\kern 0pt <}\limits_{\displaystyle\sim}}}}
\def\gapprox{\mathrel{\smash{\mathop{\kern 0pt >}\limits_{\displaystyle\sim}}}}
\begin{document}

\title{The Ratio of Gluon Distributions in Sn and C\thanks{Supported by 
the Federal Ministry of Education, Science, Research and Technology (BMBF) 
under grant no. 06~HD~742}}
\author{T. Gousset and H.J. Pirner\\
Institut f\H ur Theoretische Physik, Universit\H at Heidelberg\\
Philosophenweg 19, D-69120 Heidelberg, Germany}
\date{}

\maketitle

\vfill
\begin{abstract}
We calculate the ratio of gluon densities, $G^{\rm Sn}(x)/G^{\rm C}(x)$, for 
$0.01<x<0.1$, from the new high statistics data on 
$F_2^{\rm Sn}/F_2^{\rm C}$ taken by the NM Collaboration. For small $x$, 
the shadowing in the gluon distribution is about equal to the shadowing of 
quark distribution. The antishadowing in the gluon distribution, however, is 
roughly 10\%. We also compare with the ratio difference 
$R=\sigma_L/\sigma_T$ from Sn and C.
\end{abstract}
\vfill
\hbox to \hsize{\hfill HD-TVP-96-1}\vglue 5mm
\newpage

The investigation of nuclei with deep inelastic lepton scattering has produced 
very accurate data on nuclear structure functions $F_2^{\rm A}(x)$. 
Especially the measurement of the ratio of tin to carbon structure functions 
has been carried out recently with very high statistics~\cite{nmc}. It shows 
shadowing for $x\leq0.05$. Antishadowing ($\leq2$~\%) is visible around $x=0.1$, 
i.e. at distance scales of $r=1/xM_N\approx2\,$fm corresponding to the average 
distance between nucleons. In the interval $0.15\leq x\leq0.8$ the 
nuclear valence quark density is reduced. By measuring three different 
$\mu$-beam energies (120, 200 and 280~GeV) NMC has also been able to 
determine the weak $\ln Q^2$ dependence of the structure function ratio. The 
aim of this note is to use the leading order evolution equations to estimate the 
little known underlying gluon densities. 

There are older data~\cite{nmc92} from inelastic $J/\psi$ production at 
$z\leq 0.85$, $p_\bot^2\geq 0.4\,$(GeV/$c)^2$, which determine
$G^{\rm Sn}(x)/G^{\rm C}(x)\approx 1.13\pm0.08$ in the  interval 
$0.05\leq x\leq 0.15$. Hadronic production of $J/\psi$, $\psi'$ and $\Upsilon$ 
in $pA$ or $\pi A$ collisions~\cite{ald91} has been analysed with the gluon-gluon 
fusion process to get $G^{\rm A}(x_2)/G^p(x_2)$ in the region where the nuclear 
gluon momentum fraction $x_2$ is small~\cite{gup92,kha95}. In the small $x_2$ 
region, however, the second gluon is as soft as the third gluon which is necessary 
to form the $J/\psi$ $^3S_1$-state. Therefore one expects final state interaction 
effects in the small $x_2$ region~\cite{huf95}.

Currently there is a discussion to use nuclear beams in HERA. The study of the gluon
distribution in nuclei at small $x$ would be one of the outstanding new
opportunities to investigate nuclei on the parton level. Also the forthcoming heavy
ion experiments at RHIC and LHC need a good understanding of the nuclear parton
distributions to calculate the cross sections for hard processes initiated by
nuclear collisions.

Measurements of the gluon distribution in nuclei give experimental windows 
viewing the partonic structure of nuclear binding. Very little is known about 
the role gauge fields play in the nucleus. This has been studied for the case of 
abelian QED in a model where the hydrogene atom replaces the nucleon. In this 
model, one computes the photon density of the hydrogene molecule, 
$G^{{\rm H}_2}(x)$, and compares it to the free atom photon density, 
$G^{\rm H}(x)$. The electron orbits of the hydrogene atoms in the molecule are 
polarized and modified by the electron exchange interaction leading to a 
suppression of photons at small $x$. At the momentum corresponding to the 
relative distance of the two protons a small antishadowing peak is 
visible~\cite{pir93}. In analogy, gluon antishadowing may indicate the relative 
distance $(xM_N)^{-1}\approx 2\,$fm between the centers of nucleons, which 
act as a source for color fields. A covalent binding of quarks may manifest 
itself as a density dependent lack of long range gluons at $x<0.1$ similarly 
to the deformation of the photon cloud in the hydrogene molecule. In addition, 
in non-abelian QCD one expects at small $x$ processes where gluons from 
different nucleons overlap and merge~\cite{mue86}. Both effects have also an 
interpretation in the nuclear rest frame in terms of absorption of various 
hadronic components of the photon in the nucleus.

The strong theoretical interest, the excellent new data of NMC and the starting
discussion of HERA with nuclei motivated us to extract from the available data the
maximum information on the gluon distributions in nuclei. 
\bigskip

As usual, we define the Bjorken variable relative to nucleon kinematics
$$
x={Q^2\over 2 M_N\nu},
$$
with $M_N$ the nucleon mass, $\nu$ the virtual photon energy in the laboratory 
and $Q^2=-q^2$ its virtuality. The Leading-Log (LL) approximation reads (for a 
review see Ref.~\cite{roberts})
\begin{eqnarray}
F_2(x,Q^2)&=&x\sum_i e_i^2 q_i(x,Q^2),\\
\label{evolution}
{\partial F_2\over\partial \ln Q^2}&=&{\alpha_S\over 2\pi}
x\left(\sum_i e_i^2P_{qG}*G+P_{qq}*{F_2\over x}\right),
\end{eqnarray}
where $q_i$ and $G$ are the quark and gluon densities in the nucleon or in the 
nucleus (in the nuclear case, we shall consider distributions per nucleon). The 
splitting functions are denoted by $P_{qq}$ and $P_{qG}$. The sum runs over 
active flavors of quarks and antiquarks and $e_i$ is the corresponding charge. 
It is understood that distributions, structure functions, as well as $\alpha_S$ 
are $Q^2$ dependent.  As we will see, for $x$ sufficiently small, the second term 
in the evolution equation of $F_2$ is small in comparison with the first. 
Furthermore, the first term can be approximated in a simple way to easily 
extract the gluon distribution from the evolution of $F_2$~\cite{pry93}. In the 
nucleon case, the LL approximation becomes insufficient at very small $x$. 
Next-to-LL (NLL) corrections are about $10-20$\% at $x=0.01$. For our 
purpose of extracting ratios in the range $10^{-2}-10^{-1}$, however, the 
LL approximation is accurate enough. In the nuclear case, additionnal merging 
terms are expected to play an important role in the evolution equation with the 
increase of the gluon density. These non-linear effects are expected to appear 
at very small $x$ ($<10^{-2}$)~\cite{mue86}.
\medskip

Let us consider the first term in Eq.~(\ref{evolution}). It can be written in 
detail as
\begin{equation}
x\sum_i e_i^2\left(P_{qG}*G\right)={x\over 2}\sum_i e_i^2\int_{x/A}^1{dy\over y}
(y^2+(1-y)^2)\,G(x/y).
\end{equation}
We can replace the lower bound of the integral, $x/A$, by $x/A\to 0$, if in 
addition we impose $G(u)\equiv 0$ in the unphysical region, $u\ge A$. Because 
the gluon density is completly suppressed for $x\gapprox0.5$ this replacement 
is an accurate approximation in the small $x$ region for any reasonable ansatz 
for $G(x)$. We then expand 
$$
g\left({x\over y}\right)\equiv{x\over y}G\left({x\over y}\right)
$$
around $y=1/2$,
$$
g\left({x\over y}\right)=g(2x)-4xg'(2x)\left(y-{1\over2}\right)+\Big[(4x)^2g''(2x)
+16xg'(2x)\Big]{(y-{1\over2})^2\over 2!}+\cdots
$$
Carrying out the integration, odd terms in the expansion do not contribute and one gets 
\begin{equation}
\label{expansion}
\int_0^1 dy(y^2+(1-y)^2)g(x/y)={2\over 3}\,g(2x)
+{2\over 15}\Big[(2x)^2g''(2x)+2(2x)g'(2x)\Big]+\cdots
\end{equation}
As found in~\cite{pry93}, this expansion is efficient for standard gluon distributions 
and one can approximate
\begin{equation}
\label{approx}
\int_0^1 dy(y^2+(1-y)^2)g(x/y)\approx{2\over 3}\,g(2x).
\end{equation}
One can effectively check that for $g(u)=K\,u^{-\Delta}$, the right hand side of 
Eq.~(\ref{expansion}) becomes
$$
{2\over 3}\,K\,(2x)^{-\Delta}\left(1+{\Delta(\Delta-1)\over 5}+\cdots\right),
$$
showing that Eq.~(\ref{approx}) is accurate for $0<\Delta<1$, with an error smaller 
than 5\%.
\medskip

\begin{figure}
$$
\setlength{\unitlength}{0.240900pt}
\ifx\plotpoint\undefined\newsavebox{\plotpoint}\fi
\begin{picture}(1500,900)(0,0)
\font\gnuplot=cmr10 at 10pt
\gnuplot
\sbox{\plotpoint}{\rule[-0.200pt]{0.400pt}{0.400pt}}%
\put(176.0,113.0){\rule[-0.200pt]{4.818pt}{0.400pt}}
\put(154,113){\makebox(0,0)[r]{0.01}}
\put(1416.0,113.0){\rule[-0.200pt]{4.818pt}{0.400pt}}
\put(176.0,206.0){\rule[-0.200pt]{2.409pt}{0.400pt}}
\put(1426.0,206.0){\rule[-0.200pt]{2.409pt}{0.400pt}}
\put(176.0,260.0){\rule[-0.200pt]{2.409pt}{0.400pt}}
\put(1426.0,260.0){\rule[-0.200pt]{2.409pt}{0.400pt}}
\put(176.0,299.0){\rule[-0.200pt]{2.409pt}{0.400pt}}
\put(1426.0,299.0){\rule[-0.200pt]{2.409pt}{0.400pt}}
\put(176.0,329.0){\rule[-0.200pt]{2.409pt}{0.400pt}}
\put(1426.0,329.0){\rule[-0.200pt]{2.409pt}{0.400pt}}
\put(176.0,353.0){\rule[-0.200pt]{2.409pt}{0.400pt}}
\put(1426.0,353.0){\rule[-0.200pt]{2.409pt}{0.400pt}}
\put(176.0,374.0){\rule[-0.200pt]{2.409pt}{0.400pt}}
\put(1426.0,374.0){\rule[-0.200pt]{2.409pt}{0.400pt}}
\put(176.0,392.0){\rule[-0.200pt]{2.409pt}{0.400pt}}
\put(1426.0,392.0){\rule[-0.200pt]{2.409pt}{0.400pt}}
\put(176.0,407.0){\rule[-0.200pt]{2.409pt}{0.400pt}}
\put(1426.0,407.0){\rule[-0.200pt]{2.409pt}{0.400pt}}
\put(176.0,421.0){\rule[-0.200pt]{4.818pt}{0.400pt}}
\put(154,421){\makebox(0,0)[r]{0.1}}
\put(1416.0,421.0){\rule[-0.200pt]{4.818pt}{0.400pt}}
\put(176.0,514.0){\rule[-0.200pt]{2.409pt}{0.400pt}}
\put(1426.0,514.0){\rule[-0.200pt]{2.409pt}{0.400pt}}
\put(176.0,569.0){\rule[-0.200pt]{2.409pt}{0.400pt}}
\put(1426.0,569.0){\rule[-0.200pt]{2.409pt}{0.400pt}}
\put(176.0,607.0){\rule[-0.200pt]{2.409pt}{0.400pt}}
\put(1426.0,607.0){\rule[-0.200pt]{2.409pt}{0.400pt}}
\put(176.0,637.0){\rule[-0.200pt]{2.409pt}{0.400pt}}
\put(1426.0,637.0){\rule[-0.200pt]{2.409pt}{0.400pt}}
\put(176.0,661.0){\rule[-0.200pt]{2.409pt}{0.400pt}}
\put(1426.0,661.0){\rule[-0.200pt]{2.409pt}{0.400pt}}
\put(176.0,682.0){\rule[-0.200pt]{2.409pt}{0.400pt}}
\put(1426.0,682.0){\rule[-0.200pt]{2.409pt}{0.400pt}}
\put(176.0,700.0){\rule[-0.200pt]{2.409pt}{0.400pt}}
\put(1426.0,700.0){\rule[-0.200pt]{2.409pt}{0.400pt}}
\put(176.0,716.0){\rule[-0.200pt]{2.409pt}{0.400pt}}
\put(1426.0,716.0){\rule[-0.200pt]{2.409pt}{0.400pt}}
\put(176.0,730.0){\rule[-0.200pt]{4.818pt}{0.400pt}}
\put(154,730){\makebox(0,0)[r]{1}}
\put(1416.0,730.0){\rule[-0.200pt]{4.818pt}{0.400pt}}
\put(176.0,823.0){\rule[-0.200pt]{2.409pt}{0.400pt}}
\put(1426.0,823.0){\rule[-0.200pt]{2.409pt}{0.400pt}}
\put(176.0,877.0){\rule[-0.200pt]{2.409pt}{0.400pt}}
\put(1426.0,877.0){\rule[-0.200pt]{2.409pt}{0.400pt}}
\put(176.0,113.0){\rule[-0.200pt]{0.400pt}{4.818pt}}
\put(176,68){\makebox(0,0){0.02}}
\put(176.0,857.0){\rule[-0.200pt]{0.400pt}{4.818pt}}
\put(365.0,113.0){\rule[-0.200pt]{0.400pt}{4.818pt}}
\put(365.0,857.0){\rule[-0.200pt]{0.400pt}{4.818pt}}
\put(499.0,113.0){\rule[-0.200pt]{0.400pt}{4.818pt}}
\put(499.0,857.0){\rule[-0.200pt]{0.400pt}{4.818pt}}
\put(602.0,113.0){\rule[-0.200pt]{0.400pt}{4.818pt}}
\put(602,68){\makebox(0,0){0.05}}
\put(602.0,857.0){\rule[-0.200pt]{0.400pt}{4.818pt}}
\put(687.0,113.0){\rule[-0.200pt]{0.400pt}{4.818pt}}
\put(687.0,857.0){\rule[-0.200pt]{0.400pt}{4.818pt}}
\put(759.0,113.0){\rule[-0.200pt]{0.400pt}{4.818pt}}
\put(759.0,857.0){\rule[-0.200pt]{0.400pt}{4.818pt}}
\put(821.0,113.0){\rule[-0.200pt]{0.400pt}{4.818pt}}
\put(821.0,857.0){\rule[-0.200pt]{0.400pt}{4.818pt}}
\put(876.0,113.0){\rule[-0.200pt]{0.400pt}{4.818pt}}
\put(876.0,857.0){\rule[-0.200pt]{0.400pt}{4.818pt}}
\put(925.0,113.0){\rule[-0.200pt]{0.400pt}{4.818pt}}
\put(925,68){\makebox(0,0){0.1}}
\put(925.0,857.0){\rule[-0.200pt]{0.400pt}{4.818pt}}
\put(1247.0,113.0){\rule[-0.200pt]{0.400pt}{4.818pt}}
\put(1247,68){\makebox(0,0){0.2}}
\put(1247.0,857.0){\rule[-0.200pt]{0.400pt}{4.818pt}}
\put(176.0,113.0){\rule[-0.200pt]{303.534pt}{0.400pt}}
\put(1436.0,113.0){\rule[-0.200pt]{0.400pt}{184.048pt}}
\put(176.0,877.0){\rule[-0.200pt]{303.534pt}{0.400pt}}
\put(1500,68){\makebox(0,0){$x$}}
\put(365,206){\makebox(0,0)[l]{2 GeV$^2$}}
\put(365,407){\makebox(0,0)[l]{7 GeV$^2$}}
\put(365,514){\makebox(0,0)[l]{12 GeV$^2$}}
\put(176.0,113.0){\rule[-0.200pt]{0.400pt}{184.048pt}}
\put(1306,812){\makebox(0,0)[r]{$-\Delta_{Q^2}(x)$}}
\put(1328.0,812.0){\rule[-0.200pt]{15.899pt}{0.400pt}}
\put(176,210){\usebox{\plotpoint}}
\multiput(176.00,210.58)(2.888,0.497){63}{\rule{2.391pt}{0.120pt}}
\multiput(176.00,209.17)(184.038,33.000){2}{\rule{1.195pt}{0.400pt}}
\multiput(365.00,243.58)(2.253,0.497){57}{\rule{1.887pt}{0.120pt}}
\multiput(365.00,242.17)(130.084,30.000){2}{\rule{0.943pt}{0.400pt}}
\multiput(499.00,273.58)(2.081,0.497){47}{\rule{1.748pt}{0.120pt}}
\multiput(499.00,272.17)(99.372,25.000){2}{\rule{0.874pt}{0.400pt}}
\multiput(602.00,298.58)(2.154,0.496){37}{\rule{1.800pt}{0.119pt}}
\multiput(602.00,297.17)(81.264,20.000){2}{\rule{0.900pt}{0.400pt}}
\multiput(687.00,318.58)(2.152,0.495){31}{\rule{1.794pt}{0.119pt}}
\multiput(687.00,317.17)(68.276,17.000){2}{\rule{0.897pt}{0.400pt}}
\multiput(759.00,335.58)(2.105,0.494){27}{\rule{1.753pt}{0.119pt}}
\multiput(759.00,334.17)(58.361,15.000){2}{\rule{0.877pt}{0.400pt}}
\multiput(821.00,350.58)(2.003,0.494){25}{\rule{1.671pt}{0.119pt}}
\multiput(821.00,349.17)(51.531,14.000){2}{\rule{0.836pt}{0.400pt}}
\multiput(876.00,364.58)(1.782,0.494){25}{\rule{1.500pt}{0.119pt}}
\multiput(876.00,363.17)(45.887,14.000){2}{\rule{0.750pt}{0.400pt}}
\multiput(925.00,378.58)(1.726,0.493){23}{\rule{1.454pt}{0.119pt}}
\multiput(925.00,377.17)(40.982,13.000){2}{\rule{0.727pt}{0.400pt}}
\multiput(969.00,391.58)(1.607,0.493){23}{\rule{1.362pt}{0.119pt}}
\multiput(969.00,390.17)(38.174,13.000){2}{\rule{0.681pt}{0.400pt}}
\multiput(1010.00,404.58)(1.573,0.492){21}{\rule{1.333pt}{0.119pt}}
\multiput(1010.00,403.17)(34.233,12.000){2}{\rule{0.667pt}{0.400pt}}
\multiput(1047.00,416.58)(1.444,0.492){21}{\rule{1.233pt}{0.119pt}}
\multiput(1047.00,415.17)(31.440,12.000){2}{\rule{0.617pt}{0.400pt}}
\multiput(1081.00,428.58)(1.486,0.492){19}{\rule{1.264pt}{0.118pt}}
\multiput(1081.00,427.17)(29.377,11.000){2}{\rule{0.632pt}{0.400pt}}
\multiput(1113.00,439.58)(1.439,0.492){19}{\rule{1.227pt}{0.118pt}}
\multiput(1113.00,438.17)(28.453,11.000){2}{\rule{0.614pt}{0.400pt}}
\multiput(1144.00,450.58)(1.433,0.491){17}{\rule{1.220pt}{0.118pt}}
\multiput(1144.00,449.17)(25.468,10.000){2}{\rule{0.610pt}{0.400pt}}
\multiput(1172.00,460.59)(1.485,0.489){15}{\rule{1.256pt}{0.118pt}}
\multiput(1172.00,459.17)(23.394,9.000){2}{\rule{0.628pt}{0.400pt}}
\multiput(1198.00,469.59)(1.616,0.488){13}{\rule{1.350pt}{0.117pt}}
\multiput(1198.00,468.17)(22.198,8.000){2}{\rule{0.675pt}{0.400pt}}
\multiput(1223.00,477.59)(1.550,0.488){13}{\rule{1.300pt}{0.117pt}}
\multiput(1223.00,476.17)(21.302,8.000){2}{\rule{0.650pt}{0.400pt}}
\multiput(1247.00,485.59)(1.713,0.485){11}{\rule{1.414pt}{0.117pt}}
\multiput(1247.00,484.17)(20.065,7.000){2}{\rule{0.707pt}{0.400pt}}
\multiput(1270.00,492.59)(1.637,0.485){11}{\rule{1.357pt}{0.117pt}}
\multiput(1270.00,491.17)(19.183,7.000){2}{\rule{0.679pt}{0.400pt}}
\multiput(1292.00,499.59)(2.157,0.477){7}{\rule{1.700pt}{0.115pt}}
\multiput(1292.00,498.17)(16.472,5.000){2}{\rule{0.850pt}{0.400pt}}
\multiput(1312.00,504.59)(2.157,0.477){7}{\rule{1.700pt}{0.115pt}}
\multiput(1312.00,503.17)(16.472,5.000){2}{\rule{0.850pt}{0.400pt}}
\multiput(1332.00,509.59)(2.046,0.477){7}{\rule{1.620pt}{0.115pt}}
\multiput(1332.00,508.17)(15.638,5.000){2}{\rule{0.810pt}{0.400pt}}
\multiput(1351.00,514.61)(3.811,0.447){3}{\rule{2.500pt}{0.108pt}}
\multiput(1351.00,513.17)(12.811,3.000){2}{\rule{1.250pt}{0.400pt}}
\multiput(1369.00,517.60)(2.528,0.468){5}{\rule{1.900pt}{0.113pt}}
\multiput(1369.00,516.17)(14.056,4.000){2}{\rule{0.950pt}{0.400pt}}
\put(1387,521.17){\rule{3.500pt}{0.400pt}}
\multiput(1387.00,520.17)(9.736,2.000){2}{\rule{1.750pt}{0.400pt}}
\put(1404,523.17){\rule{3.300pt}{0.400pt}}
\multiput(1404.00,522.17)(9.151,2.000){2}{\rule{1.650pt}{0.400pt}}
\put(1420,524.67){\rule{3.854pt}{0.400pt}}
\multiput(1420.00,524.17)(8.000,1.000){2}{\rule{1.927pt}{0.400pt}}
\put(176,431){\usebox{\plotpoint}}
\multiput(176.00,431.58)(9.825,0.491){17}{\rule{7.660pt}{0.118pt}}
\multiput(176.00,430.17)(173.101,10.000){2}{\rule{3.830pt}{0.400pt}}
\multiput(365.00,441.59)(8.814,0.488){13}{\rule{6.800pt}{0.117pt}}
\multiput(365.00,440.17)(119.886,8.000){2}{\rule{3.400pt}{0.400pt}}
\multiput(499.00,449.59)(7.815,0.485){11}{\rule{5.986pt}{0.117pt}}
\multiput(499.00,448.17)(90.576,7.000){2}{\rule{2.993pt}{0.400pt}}
\multiput(602.00,456.59)(7.633,0.482){9}{\rule{5.767pt}{0.116pt}}
\multiput(602.00,455.17)(73.031,6.000){2}{\rule{2.883pt}{0.400pt}}
\multiput(687.00,462.59)(6.457,0.482){9}{\rule{4.900pt}{0.116pt}}
\multiput(687.00,461.17)(61.830,6.000){2}{\rule{2.450pt}{0.400pt}}
\multiput(759.00,468.59)(5.553,0.482){9}{\rule{4.233pt}{0.116pt}}
\multiput(759.00,467.17)(53.214,6.000){2}{\rule{2.117pt}{0.400pt}}
\multiput(821.00,474.59)(6.053,0.477){7}{\rule{4.500pt}{0.115pt}}
\multiput(821.00,473.17)(45.660,5.000){2}{\rule{2.250pt}{0.400pt}}
\multiput(876.00,479.59)(5.385,0.477){7}{\rule{4.020pt}{0.115pt}}
\multiput(876.00,478.17)(40.656,5.000){2}{\rule{2.010pt}{0.400pt}}
\multiput(925.00,484.59)(3.926,0.482){9}{\rule{3.033pt}{0.116pt}}
\multiput(925.00,483.17)(37.704,6.000){2}{\rule{1.517pt}{0.400pt}}
\multiput(969.00,490.59)(4.495,0.477){7}{\rule{3.380pt}{0.115pt}}
\multiput(969.00,489.17)(33.985,5.000){2}{\rule{1.690pt}{0.400pt}}
\multiput(1010.00,495.59)(3.293,0.482){9}{\rule{2.567pt}{0.116pt}}
\multiput(1010.00,494.17)(31.673,6.000){2}{\rule{1.283pt}{0.400pt}}
\multiput(1047.00,501.59)(3.716,0.477){7}{\rule{2.820pt}{0.115pt}}
\multiput(1047.00,500.17)(28.147,5.000){2}{\rule{1.410pt}{0.400pt}}
\multiput(1081.00,506.59)(3.493,0.477){7}{\rule{2.660pt}{0.115pt}}
\multiput(1081.00,505.17)(26.479,5.000){2}{\rule{1.330pt}{0.400pt}}
\multiput(1113.00,511.59)(3.382,0.477){7}{\rule{2.580pt}{0.115pt}}
\multiput(1113.00,510.17)(25.645,5.000){2}{\rule{1.290pt}{0.400pt}}
\multiput(1144.00,516.59)(3.048,0.477){7}{\rule{2.340pt}{0.115pt}}
\multiput(1144.00,515.17)(23.143,5.000){2}{\rule{1.170pt}{0.400pt}}
\multiput(1172.00,521.59)(2.825,0.477){7}{\rule{2.180pt}{0.115pt}}
\multiput(1172.00,520.17)(21.475,5.000){2}{\rule{1.090pt}{0.400pt}}
\multiput(1198.00,526.60)(3.552,0.468){5}{\rule{2.600pt}{0.113pt}}
\multiput(1198.00,525.17)(19.604,4.000){2}{\rule{1.300pt}{0.400pt}}
\multiput(1223.00,530.60)(3.406,0.468){5}{\rule{2.500pt}{0.113pt}}
\multiput(1223.00,529.17)(18.811,4.000){2}{\rule{1.250pt}{0.400pt}}
\multiput(1247.00,534.60)(3.259,0.468){5}{\rule{2.400pt}{0.113pt}}
\multiput(1247.00,533.17)(18.019,4.000){2}{\rule{1.200pt}{0.400pt}}
\multiput(1270.00,538.61)(4.704,0.447){3}{\rule{3.033pt}{0.108pt}}
\multiput(1270.00,537.17)(15.704,3.000){2}{\rule{1.517pt}{0.400pt}}
\multiput(1292.00,541.61)(4.258,0.447){3}{\rule{2.767pt}{0.108pt}}
\multiput(1292.00,540.17)(14.258,3.000){2}{\rule{1.383pt}{0.400pt}}
\multiput(1312.00,544.61)(4.258,0.447){3}{\rule{2.767pt}{0.108pt}}
\multiput(1312.00,543.17)(14.258,3.000){2}{\rule{1.383pt}{0.400pt}}
\put(1332,547.17){\rule{3.900pt}{0.400pt}}
\multiput(1332.00,546.17)(10.905,2.000){2}{\rule{1.950pt}{0.400pt}}
\multiput(1351.00,549.61)(3.811,0.447){3}{\rule{2.500pt}{0.108pt}}
\multiput(1351.00,548.17)(12.811,3.000){2}{\rule{1.250pt}{0.400pt}}
\put(1369,551.67){\rule{4.336pt}{0.400pt}}
\multiput(1369.00,551.17)(9.000,1.000){2}{\rule{2.168pt}{0.400pt}}
\put(1387,553.17){\rule{3.500pt}{0.400pt}}
\multiput(1387.00,552.17)(9.736,2.000){2}{\rule{1.750pt}{0.400pt}}
\put(1404,554.67){\rule{3.854pt}{0.400pt}}
\multiput(1404.00,554.17)(8.000,1.000){2}{\rule{1.927pt}{0.400pt}}
\put(1420.0,556.0){\rule[-0.200pt]{3.854pt}{0.400pt}}
\put(176,473){\usebox{\plotpoint}}
\multiput(176.00,473.59)(14.375,0.485){11}{\rule{10.900pt}{0.117pt}}
\multiput(176.00,472.17)(166.377,7.000){2}{\rule{5.450pt}{0.400pt}}
\multiput(365.00,480.59)(14.848,0.477){7}{\rule{10.820pt}{0.115pt}}
\multiput(365.00,479.17)(111.543,5.000){2}{\rule{5.410pt}{0.400pt}}
\multiput(499.00,485.59)(11.397,0.477){7}{\rule{8.340pt}{0.115pt}}
\multiput(499.00,484.17)(85.690,5.000){2}{\rule{4.170pt}{0.400pt}}
\multiput(602.00,490.60)(12.325,0.468){5}{\rule{8.600pt}{0.113pt}}
\multiput(602.00,489.17)(67.150,4.000){2}{\rule{4.300pt}{0.400pt}}
\multiput(687.00,494.60)(10.424,0.468){5}{\rule{7.300pt}{0.113pt}}
\multiput(687.00,493.17)(56.848,4.000){2}{\rule{3.650pt}{0.400pt}}
\multiput(759.00,498.60)(8.962,0.468){5}{\rule{6.300pt}{0.113pt}}
\multiput(759.00,497.17)(48.924,4.000){2}{\rule{3.150pt}{0.400pt}}
\multiput(821.00,502.60)(7.938,0.468){5}{\rule{5.600pt}{0.113pt}}
\multiput(821.00,501.17)(43.377,4.000){2}{\rule{2.800pt}{0.400pt}}
\multiput(876.00,506.60)(7.061,0.468){5}{\rule{5.000pt}{0.113pt}}
\multiput(876.00,505.17)(38.622,4.000){2}{\rule{2.500pt}{0.400pt}}
\multiput(925.00,510.60)(6.330,0.468){5}{\rule{4.500pt}{0.113pt}}
\multiput(925.00,509.17)(34.660,4.000){2}{\rule{2.250pt}{0.400pt}}
\multiput(969.00,514.60)(5.891,0.468){5}{\rule{4.200pt}{0.113pt}}
\multiput(969.00,513.17)(32.283,4.000){2}{\rule{2.100pt}{0.400pt}}
\multiput(1010.00,518.60)(5.306,0.468){5}{\rule{3.800pt}{0.113pt}}
\multiput(1010.00,517.17)(29.113,4.000){2}{\rule{1.900pt}{0.400pt}}
\multiput(1047.00,522.60)(4.868,0.468){5}{\rule{3.500pt}{0.113pt}}
\multiput(1047.00,521.17)(26.736,4.000){2}{\rule{1.750pt}{0.400pt}}
\multiput(1081.00,526.60)(4.575,0.468){5}{\rule{3.300pt}{0.113pt}}
\multiput(1081.00,525.17)(25.151,4.000){2}{\rule{1.650pt}{0.400pt}}
\multiput(1113.00,530.60)(4.429,0.468){5}{\rule{3.200pt}{0.113pt}}
\multiput(1113.00,529.17)(24.358,4.000){2}{\rule{1.600pt}{0.400pt}}
\multiput(1144.00,534.61)(6.044,0.447){3}{\rule{3.833pt}{0.108pt}}
\multiput(1144.00,533.17)(20.044,3.000){2}{\rule{1.917pt}{0.400pt}}
\multiput(1172.00,537.60)(3.698,0.468){5}{\rule{2.700pt}{0.113pt}}
\multiput(1172.00,536.17)(20.396,4.000){2}{\rule{1.350pt}{0.400pt}}
\multiput(1198.00,541.61)(5.374,0.447){3}{\rule{3.433pt}{0.108pt}}
\multiput(1198.00,540.17)(17.874,3.000){2}{\rule{1.717pt}{0.400pt}}
\multiput(1223.00,544.61)(5.151,0.447){3}{\rule{3.300pt}{0.108pt}}
\multiput(1223.00,543.17)(17.151,3.000){2}{\rule{1.650pt}{0.400pt}}
\multiput(1247.00,547.61)(4.927,0.447){3}{\rule{3.167pt}{0.108pt}}
\multiput(1247.00,546.17)(16.427,3.000){2}{\rule{1.583pt}{0.400pt}}
\put(1270,550.17){\rule{4.500pt}{0.400pt}}
\multiput(1270.00,549.17)(12.660,2.000){2}{\rule{2.250pt}{0.400pt}}
\multiput(1292.00,552.61)(4.258,0.447){3}{\rule{2.767pt}{0.108pt}}
\multiput(1292.00,551.17)(14.258,3.000){2}{\rule{1.383pt}{0.400pt}}
\put(1312,554.67){\rule{4.818pt}{0.400pt}}
\multiput(1312.00,554.17)(10.000,1.000){2}{\rule{2.409pt}{0.400pt}}
\put(1332,556.17){\rule{3.900pt}{0.400pt}}
\multiput(1332.00,555.17)(10.905,2.000){2}{\rule{1.950pt}{0.400pt}}
\put(1351,557.67){\rule{4.336pt}{0.400pt}}
\multiput(1351.00,557.17)(9.000,1.000){2}{\rule{2.168pt}{0.400pt}}
\put(1369,558.67){\rule{4.336pt}{0.400pt}}
\multiput(1369.00,558.17)(9.000,1.000){2}{\rule{2.168pt}{0.400pt}}
\put(1387,559.67){\rule{4.095pt}{0.400pt}}
\multiput(1387.00,559.17)(8.500,1.000){2}{\rule{2.048pt}{0.400pt}}
\put(1404,560.67){\rule{3.854pt}{0.400pt}}
\multiput(1404.00,560.17)(8.000,1.000){2}{\rule{1.927pt}{0.400pt}}
\put(1420.0,562.0){\rule[-0.200pt]{3.854pt}{0.400pt}}
\sbox{\plotpoint}{\rule[-0.500pt]{1.000pt}{1.000pt}}%
\put(1306,732){\makebox(0,0)[r]{$2x\,G(2x)$}}
\multiput(1328,732)(20.756,0.000){4}{\usebox{\plotpoint}}
\put(176,870){\usebox{\plotpoint}}
\put(176.00,870.00){\usebox{\plotpoint}}
\put(196.69,868.36){\usebox{\plotpoint}}
\multiput(201,868)(20.694,-1.592){0}{\usebox{\plotpoint}}
\put(217.38,866.74){\usebox{\plotpoint}}
\put(238.08,865.15){\usebox{\plotpoint}}
\multiput(240,865)(20.684,-1.724){0}{\usebox{\plotpoint}}
\put(258.77,863.48){\usebox{\plotpoint}}
\multiput(265,863)(20.514,-3.156){0}{\usebox{\plotpoint}}
\put(279.35,860.90){\usebox{\plotpoint}}
\put(300.04,859.25){\usebox{\plotpoint}}
\multiput(303,859)(20.514,-3.156){0}{\usebox{\plotpoint}}
\put(320.61,856.65){\usebox{\plotpoint}}
\multiput(329,856)(20.684,-1.724){0}{\usebox{\plotpoint}}
\put(341.30,854.95){\usebox{\plotpoint}}
\put(361.88,852.39){\usebox{\plotpoint}}
\multiput(367,852)(20.514,-3.156){0}{\usebox{\plotpoint}}
\put(382.46,849.79){\usebox{\plotpoint}}
\put(403.05,847.30){\usebox{\plotpoint}}
\multiput(405,847)(20.514,-3.156){0}{\usebox{\plotpoint}}
\put(423.62,844.57){\usebox{\plotpoint}}
\multiput(431,844)(20.473,-3.412){0}{\usebox{\plotpoint}}
\put(444.17,841.82){\usebox{\plotpoint}}
\put(464.69,838.66){\usebox{\plotpoint}}
\multiput(469,838)(20.473,-3.412){0}{\usebox{\plotpoint}}
\put(485.18,835.36){\usebox{\plotpoint}}
\put(505.69,832.20){\usebox{\plotpoint}}
\multiput(507,832)(20.514,-3.156){0}{\usebox{\plotpoint}}
\put(526.19,828.97){\usebox{\plotpoint}}
\multiput(532,828)(20.514,-3.156){0}{\usebox{\plotpoint}}
\put(546.67,825.61){\usebox{\plotpoint}}
\put(567.02,821.61){\usebox{\plotpoint}}
\multiput(571,821)(20.473,-3.412){0}{\usebox{\plotpoint}}
\put(587.45,817.97){\usebox{\plotpoint}}
\put(607.67,813.31){\usebox{\plotpoint}}
\multiput(609,813)(20.473,-3.412){0}{\usebox{\plotpoint}}
\put(628.04,809.38){\usebox{\plotpoint}}
\multiput(634,808)(20.224,-4.667){0}{\usebox{\plotpoint}}
\put(648.27,804.71){\usebox{\plotpoint}}
\put(668.45,799.89){\usebox{\plotpoint}}
\multiput(672,799)(20.224,-4.667){0}{\usebox{\plotpoint}}
\put(688.66,795.16){\usebox{\plotpoint}}
\put(708.68,789.71){\usebox{\plotpoint}}
\multiput(711,789)(20.136,-5.034){0}{\usebox{\plotpoint}}
\put(728.69,784.25){\usebox{\plotpoint}}
\put(748.77,779.05){\usebox{\plotpoint}}
\multiput(749,779)(19.690,-6.563){0}{\usebox{\plotpoint}}
\put(768.53,772.68){\usebox{\plotpoint}}
\multiput(774,771)(19.838,-6.104){0}{\usebox{\plotpoint}}
\put(788.36,766.58){\usebox{\plotpoint}}
\put(808.14,760.29){\usebox{\plotpoint}}
\multiput(812,759)(19.372,-7.451){0}{\usebox{\plotpoint}}
\put(827.64,753.19){\usebox{\plotpoint}}
\put(847.25,746.44){\usebox{\plotpoint}}
\multiput(851,745)(19.159,-7.983){0}{\usebox{\plotpoint}}
\put(866.49,738.66){\usebox{\plotpoint}}
\put(885.86,731.21){\usebox{\plotpoint}}
\multiput(889,730)(18.564,-9.282){0}{\usebox{\plotpoint}}
\put(904.71,722.57){\usebox{\plotpoint}}
\put(923.81,714.47){\usebox{\plotpoint}}
\multiput(927,713)(18.845,-8.698){0}{\usebox{\plotpoint}}
\put(942.62,705.69){\usebox{\plotpoint}}
\put(961.04,696.13){\usebox{\plotpoint}}
\multiput(965,694)(18.845,-8.698){0}{\usebox{\plotpoint}}
\put(979.71,687.08){\usebox{\plotpoint}}
\put(997.85,677.01){\usebox{\plotpoint}}
\put(1015.60,666.25){\usebox{\plotpoint}}
\multiput(1016,666)(17.677,-10.878){0}{\usebox{\plotpoint}}
\put(1033.17,655.22){\usebox{\plotpoint}}
\put(1050.67,644.05){\usebox{\plotpoint}}
\multiput(1054,642)(17.677,-10.878){0}{\usebox{\plotpoint}}
\put(1068.30,633.10){\usebox{\plotpoint}}
\put(1085.01,620.82){\usebox{\plotpoint}}
\put(1101.58,608.36){\usebox{\plotpoint}}
\multiput(1105,606)(16.451,-12.655){0}{\usebox{\plotpoint}}
\put(1118.16,595.88){\usebox{\plotpoint}}
\put(1134.36,582.92){\usebox{\plotpoint}}
\put(1149.64,568.87){\usebox{\plotpoint}}
\put(1165.23,555.19){\usebox{\plotpoint}}
\put(1180.19,540.81){\usebox{\plotpoint}}
\multiput(1181,540)(14.676,-14.676){0}{\usebox{\plotpoint}}
\put(1194.83,526.11){\usebox{\plotpoint}}
\put(1209.03,510.97){\usebox{\plotpoint}}
\put(1223.27,495.91){\usebox{\plotpoint}}
\put(1236.45,479.87){\usebox{\plotpoint}}
\put(1249.85,464.03){\usebox{\plotpoint}}
\put(1262.76,447.78){\usebox{\plotpoint}}
\put(1275.15,431.13){\usebox{\plotpoint}}
\put(1287.18,414.22){\usebox{\plotpoint}}
\put(1299.21,397.31){\usebox{\plotpoint}}
\multiput(1309,383)(10.298,-18.021){2}{\usebox{\plotpoint}}
\put(1331.92,344.36){\usebox{\plotpoint}}
\put(1342.55,326.53){\usebox{\plotpoint}}
\put(1352.72,308.44){\usebox{\plotpoint}}
\multiput(1360,295)(8.982,-18.712){2}{\usebox{\plotpoint}}
\put(1380.35,252.66){\usebox{\plotpoint}}
\multiput(1385,243)(8.740,-18.825){2}{\usebox{\plotpoint}}
\put(1406.22,196.03){\usebox{\plotpoint}}
\multiput(1411,185)(7.493,-19.356){2}{\usebox{\plotpoint}}
\put(1429.07,138.12){\usebox{\plotpoint}}
\end{picture}
$$
\caption{\label{delta}{\small
The correction $-\Delta(x)$ for different $Q^2$ in the nucleon case. The NMC fit
to the gluon density~\protect\cite{nmc93}, evaluated at $2x$, is also plotted for 
comparison (the $Q^2$ for this density is 7$\,$GeV$^2$).}}
\end{figure}
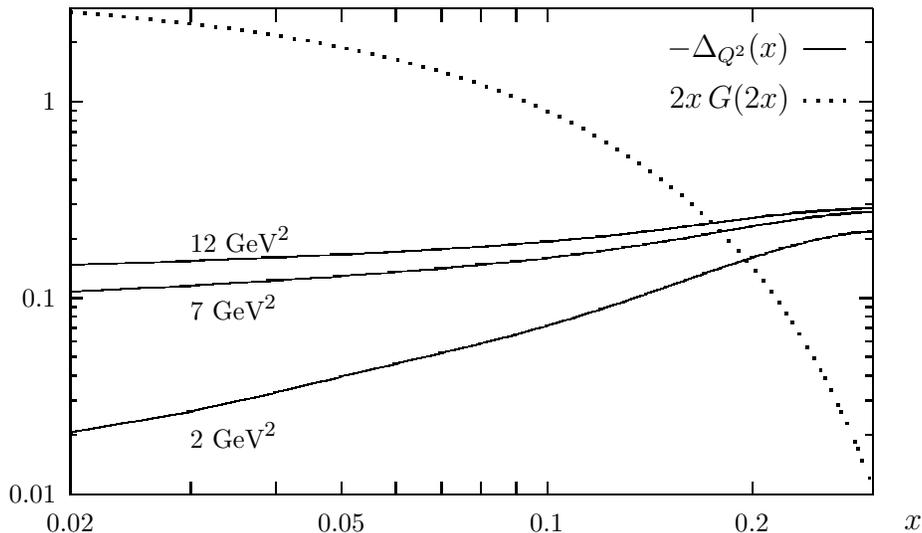

We now consider the second term in Eq.~(\ref{evolution})
\begin{equation}
\Delta(x,Q^2)=x\left(P_{qq}*{F_2\over x}\right)={4\over 3}\int_{x/A}^1 dy
\left({(1+y^2)F_2(x/y)-2F_2(x)\over 1-y}
+{3\over 2}\delta(1-y)F_2(x/y)\right);
\end{equation}
One can compute $\Delta(x,Q^2)$ from $F_2$ data. We show in Fig.~\ref{delta} 
the behavior of $\Delta$ as a function of $x$ and $Q^2$ in the case 
of the nucleon. The behavior of the gluon distribution is also plotted for 
comparison. For $x\approx 0.2$, $\Delta(x)$ is of the same order of magnitude 
as $2xG(2x)$ and indeed cancels the gluon contribution in Eq.~(\ref{evolution}) 
leading to the true scaling of $F_2$. Due to the steep rise of $G$, $\Delta$ 
becomes a small correction as soon as $x\le0.1$, thus implying a simple relation 
between $F_2$ evolution and gluon density. 
\medskip

In the following, we shall take into account both contributions to the evolution 
equation, Eq.~(\ref{evolution}), i.e. 
\goodbreak
\begin{equation}
\label{approx_evol}
{\partial F_2\over\partial \ln Q^2}\approx{\alpha_S\over 2\pi}
\left(\Delta(x,Q^2)+{2\over3}{\sum_ie_i^2\over2}2xG(2x,Q^2)\right).
\end{equation}
\medskip

We can now turn to the study of the new NMC data on tin to carbon ratio of 
structure functions. Namely, measurements of 
\begin{eqnarray}
f_1(x)&=&{F_2^{\rm Sn}\over F_2^{\rm C}},\\
f_2(x)&=&{\partial\over\partial\ln Q^2}{F_2^{\rm Sn}\over F_2^{\rm C}},
\end{eqnarray}
have been performed~\cite{nmc}. To be more precise, let us mention that the ratio, 
$f_1$, has been measured for fixed $x$ over a certain range of $Q^2$, so that the 
reported value is an average of ratios corresponding to an averaged 
$\langle Q^2\rangle$, where $\langle Q^2\rangle$ varies from one $x$ to the next. 
The evolution of ratio with $\ln Q^2$ has then been computed, giving the ``slope'', 
$f_2$, at $x$ and $\langle Q^2\rangle$. 

Using Eq.~(\ref{approx_evol}), it is easy to relate the tin to carbon ratio of gluon 
distributions 
$$
r(x,Q^2)=\left.{G^{\rm Sn}\over G^{\rm C}}\right|_{x,Q^2},
$$
to $f_1$ and $f_2$, and one finds
\begin{equation}
\label{ratio_e}
r(2x,\langle Q^2\rangle)=\left. f_1+
{f_2-{\alpha_S\over 2\pi}{\Delta^{\rm Sn}-f_1\Delta^{\rm C}\over F_2^{\rm C}}\over 
f_3-{\alpha_S\over 2\pi}{\Delta^{\rm C}\over F_2^{\rm C}}}
\right|_{x,\langle Q^2\rangle},
\end{equation}
where
\begin{equation}
\label{f_three}
f_3={\partial\ln F_2^{\rm C}\over\partial\ln Q^2}
={\partial\ln F_2^d\over\partial\ln Q^2}
+{\partial\over\partial\ln Q^2}\ln {F_2^{\rm C}\over F_2^d}.
\end{equation}
In this analysis, we use the fact that the ratio of carbon to deuteron structure 
functions shows practically no $Q^2$ dependence~\cite{nmc95}, so that we can 
drop the second term in the right hand side of Eq.~(\ref{f_three}), and use for $f_3$ 
the deuteron data of~\cite{nmc93}.   

\begin{table}[b]
$$
\begin{tabular}{cccccccccccc}
\hline\hline
$x$&0.011&0.017&0.025&0.035&0.05&0.07&0.09&0.11&0.14&0.18\\
$\langle Q^2\rangle$&1.8&2.4&3.4&4.4&5.7&7.3&8.6&9.6&11.2&12.8\\
$r$&0.87&0.88&0.98&1.03&1.06&1.08&1.09&1.07&1.09&1.05\\
$\delta r$&0.05&0.04&0.03&0.04&0.03&0.03&0.04&0.05&0.04&0.06\\
\hline\hline
\end{tabular}
$$
\caption{\label{ratio_t}{\small
The ratio of tin to carbon gluon density for different $x$ and corresponding 
$\langle Q^2\rangle$. $\delta r$ is the uncertainty on $r$ coming from the 
statistical error on the slope $f_2$.}}
\end{table}

\begin{figure}[t]
$$
\setlength{\unitlength}{0.240900pt}
\ifx\plotpoint\undefined\newsavebox{\plotpoint}\fi
\begin{picture}(1500,900)(0,0)
\font\gnuplot=cmr10 at 10pt
\gnuplot
\sbox{\plotpoint}{\rule[-0.200pt]{0.400pt}{0.400pt}}%
\put(176.0,146.0){\rule[-0.200pt]{4.818pt}{0.400pt}}
\put(154,146){\makebox(0,0)[r]{0.8}}
\put(1416.0,146.0){\rule[-0.200pt]{4.818pt}{0.400pt}}
\put(176.0,308.0){\rule[-0.200pt]{4.818pt}{0.400pt}}
\put(154,308){\makebox(0,0)[r]{0.9}}
\put(1416.0,308.0){\rule[-0.200pt]{4.818pt}{0.400pt}}
\put(176.0,471.0){\rule[-0.200pt]{4.818pt}{0.400pt}}
\put(154,471){\makebox(0,0)[r]{1}}
\put(1416.0,471.0){\rule[-0.200pt]{4.818pt}{0.400pt}}
\put(176.0,633.0){\rule[-0.200pt]{4.818pt}{0.400pt}}
\put(154,633){\makebox(0,0)[r]{1.1}}
\put(1416.0,633.0){\rule[-0.200pt]{4.818pt}{0.400pt}}
\put(176.0,796.0){\rule[-0.200pt]{4.818pt}{0.400pt}}
\put(154,796){\makebox(0,0)[r]{1.2}}
\put(1416.0,796.0){\rule[-0.200pt]{4.818pt}{0.400pt}}
\put(176.0,113.0){\rule[-0.200pt]{0.400pt}{2.409pt}}
\put(176.0,867.0){\rule[-0.200pt]{0.400pt}{2.409pt}}
\put(234.0,113.0){\rule[-0.200pt]{0.400pt}{2.409pt}}
\put(234.0,867.0){\rule[-0.200pt]{0.400pt}{2.409pt}}
\put(282.0,113.0){\rule[-0.200pt]{0.400pt}{2.409pt}}
\put(282.0,867.0){\rule[-0.200pt]{0.400pt}{2.409pt}}
\put(322.0,113.0){\rule[-0.200pt]{0.400pt}{2.409pt}}
\put(322.0,867.0){\rule[-0.200pt]{0.400pt}{2.409pt}}
\put(357.0,113.0){\rule[-0.200pt]{0.400pt}{2.409pt}}
\put(357.0,867.0){\rule[-0.200pt]{0.400pt}{2.409pt}}
\put(388.0,113.0){\rule[-0.200pt]{0.400pt}{2.409pt}}
\put(388.0,867.0){\rule[-0.200pt]{0.400pt}{2.409pt}}
\put(415.0,113.0){\rule[-0.200pt]{0.400pt}{4.818pt}}
\put(415,68){\makebox(0,0){0.01}}
\put(415.0,857.0){\rule[-0.200pt]{0.400pt}{4.818pt}}
\put(596.0,113.0){\rule[-0.200pt]{0.400pt}{2.409pt}}
\put(596.0,867.0){\rule[-0.200pt]{0.400pt}{2.409pt}}
\put(702.0,113.0){\rule[-0.200pt]{0.400pt}{2.409pt}}
\put(702.0,867.0){\rule[-0.200pt]{0.400pt}{2.409pt}}
\put(777.0,113.0){\rule[-0.200pt]{0.400pt}{2.409pt}}
\put(777.0,867.0){\rule[-0.200pt]{0.400pt}{2.409pt}}
\put(835.0,113.0){\rule[-0.200pt]{0.400pt}{2.409pt}}
\put(835.0,867.0){\rule[-0.200pt]{0.400pt}{2.409pt}}
\put(883.0,113.0){\rule[-0.200pt]{0.400pt}{2.409pt}}
\put(883.0,867.0){\rule[-0.200pt]{0.400pt}{2.409pt}}
\put(923.0,113.0){\rule[-0.200pt]{0.400pt}{2.409pt}}
\put(923.0,867.0){\rule[-0.200pt]{0.400pt}{2.409pt}}
\put(958.0,113.0){\rule[-0.200pt]{0.400pt}{2.409pt}}
\put(958.0,867.0){\rule[-0.200pt]{0.400pt}{2.409pt}}
\put(989.0,113.0){\rule[-0.200pt]{0.400pt}{2.409pt}}
\put(989.0,867.0){\rule[-0.200pt]{0.400pt}{2.409pt}}
\put(1016.0,113.0){\rule[-0.200pt]{0.400pt}{4.818pt}}
\put(1016,68){\makebox(0,0){0.1}}
\put(1016.0,857.0){\rule[-0.200pt]{0.400pt}{4.818pt}}
\put(1197.0,113.0){\rule[-0.200pt]{0.400pt}{2.409pt}}
\put(1197.0,867.0){\rule[-0.200pt]{0.400pt}{2.409pt}}
\put(1303.0,113.0){\rule[-0.200pt]{0.400pt}{2.409pt}}
\put(1303.0,867.0){\rule[-0.200pt]{0.400pt}{2.409pt}}
\put(1378.0,113.0){\rule[-0.200pt]{0.400pt}{2.409pt}}
\put(1378.0,867.0){\rule[-0.200pt]{0.400pt}{2.409pt}}
\put(1436.0,113.0){\rule[-0.200pt]{0.400pt}{2.409pt}}
\put(1436.0,867.0){\rule[-0.200pt]{0.400pt}{2.409pt}}
\put(176.0,113.0){\rule[-0.200pt]{303.534pt}{0.400pt}}
\put(1436.0,113.0){\rule[-0.200pt]{0.400pt}{184.048pt}}
\put(176.0,877.0){\rule[-0.200pt]{303.534pt}{0.400pt}}
\put(1500,68){\makebox(0,0){$x$}}
\put(176.0,113.0){\rule[-0.200pt]{0.400pt}{184.048pt}}
\put(350,812){\makebox(0,0)[r]{$f_1(x)$}}
\put(400,812){\makebox(0,0){$+$}}
\put(259,176){\makebox(0,0){$+$}}
\put(373,228){\makebox(0,0){$+$}}
\put(473,290){\makebox(0,0){$+$}}
\put(561,341){\makebox(0,0){$+$}}
\put(654,409){\makebox(0,0){$+$}}
\put(742,440){\makebox(0,0){$+$}}
\put(808,458){\makebox(0,0){$+$}}
\put(860,484){\makebox(0,0){$+$}}
\put(923,485){\makebox(0,0){$+$}}
\put(989,488){\makebox(0,0){$+$}}
\put(1074,490){\makebox(0,0){$+$}}
\put(1162,493){\makebox(0,0){$+$}}
\put(1255,480){\makebox(0,0){$+$}}
\put(1343,462){\makebox(0,0){$+$}}
\put(350,745){\makebox(0,0)[r]{$r(x)$}}
\put(400,745){\circle{18}}
\put(440,264){\circle{18}}
\put(554,269){\circle{18}}
\put(654,432){\circle{18}}
\put(742,523){\circle{18}}
\put(835,565){\circle{18}}
\put(923,609){\circle{18}}
\put(989,617){\circle{18}}
\put(1041,584){\circle{18}}
\put(1104,625){\circle{18}}
\put(1169,555){\circle{18}}
\put(440.0,188.0){\rule[-0.200pt]{0.400pt}{36.858pt}}
\put(430.0,188.0){\rule[-0.200pt]{4.818pt}{0.400pt}}
\put(430.0,341.0){\rule[-0.200pt]{4.818pt}{0.400pt}}
\put(554.0,209.0){\rule[-0.200pt]{0.400pt}{28.908pt}}
\put(544.0,209.0){\rule[-0.200pt]{4.818pt}{0.400pt}}
\put(544.0,329.0){\rule[-0.200pt]{4.818pt}{0.400pt}}
\put(654.0,386.0){\rule[-0.200pt]{0.400pt}{21.922pt}}
\put(644.0,386.0){\rule[-0.200pt]{4.818pt}{0.400pt}}
\put(644.0,477.0){\rule[-0.200pt]{4.818pt}{0.400pt}}
\put(742.0,462.0){\rule[-0.200pt]{0.400pt}{29.149pt}}
\put(732.0,462.0){\rule[-0.200pt]{4.818pt}{0.400pt}}
\put(732.0,583.0){\rule[-0.200pt]{4.818pt}{0.400pt}}
\put(835.0,519.0){\rule[-0.200pt]{0.400pt}{21.922pt}}
\put(825.0,519.0){\rule[-0.200pt]{4.818pt}{0.400pt}}
\put(825.0,610.0){\rule[-0.200pt]{4.818pt}{0.400pt}}
\put(923.0,554.0){\rule[-0.200pt]{0.400pt}{26.499pt}}
\put(913.0,554.0){\rule[-0.200pt]{4.818pt}{0.400pt}}
\put(913.0,664.0){\rule[-0.200pt]{4.818pt}{0.400pt}}
\put(989.0,549.0){\rule[-0.200pt]{0.400pt}{32.762pt}}
\put(979.0,549.0){\rule[-0.200pt]{4.818pt}{0.400pt}}
\put(979.0,685.0){\rule[-0.200pt]{4.818pt}{0.400pt}}
\put(1041.0,502.0){\rule[-0.200pt]{0.400pt}{39.748pt}}
\put(1031.0,502.0){\rule[-0.200pt]{4.818pt}{0.400pt}}
\put(1031.0,667.0){\rule[-0.200pt]{4.818pt}{0.400pt}}
\put(1104.0,557.0){\rule[-0.200pt]{0.400pt}{32.762pt}}
\put(1094.0,557.0){\rule[-0.200pt]{4.818pt}{0.400pt}}
\put(1094.0,693.0){\rule[-0.200pt]{4.818pt}{0.400pt}}
\put(1169.0,459.0){\rule[-0.200pt]{0.400pt}{46.253pt}}
\put(1159.0,459.0){\rule[-0.200pt]{4.818pt}{0.400pt}}
\put(1159.0,651.0){\rule[-0.200pt]{4.818pt}{0.400pt}}
\sbox{\plotpoint}{\rule[-0.400pt]{0.800pt}{0.800pt}}%
\put(176,471){\usebox{\plotpoint}}
\put(176.0,471.0){\rule[-0.100pt]{303.534pt}{0.200pt}}
\sbox{\plotpoint}{\rule[-0.600pt]{1.200pt}{1.200pt}}%
\put(800,682){\makebox(0,0)[r]{$J/\psi$}}
\put(835.0,552.0){\rule[-0.050pt]{.100pt}{62.634pt}}
\put(835.0,552.0){\rule[-0.050pt]{69.138pt}{.100pt}}
\put(835.0,812.0){\rule[-0.050pt]{69.138pt}{.100pt}}
\put(1122.0,552.0){\rule[-0.050pt]{.100pt}{62.734pt}}
\put(835.0,682.0){\rule[-0.050pt]{69.138pt}{.100pt}}
\end{picture}
$$
\caption{\label{ratio_f}{\small
The ratio $r(x)=G^{\rm Sn}(x)/G^{\rm C}(x)$ of tin to carbon gluon density as a 
function of $x$ together with the ratio of structure function, 
$f_1(x)=F_2^{\rm Sn}(x)/F_2^{\rm C}(x)$. The statistical error on $f_1$ is less 
than 1\% in the whole range of $x$. The box represents the extraction of $r$ 
from $J/\psi$ electroproduction data~\protect\cite{nmc92} (see text).}}
\end{figure}
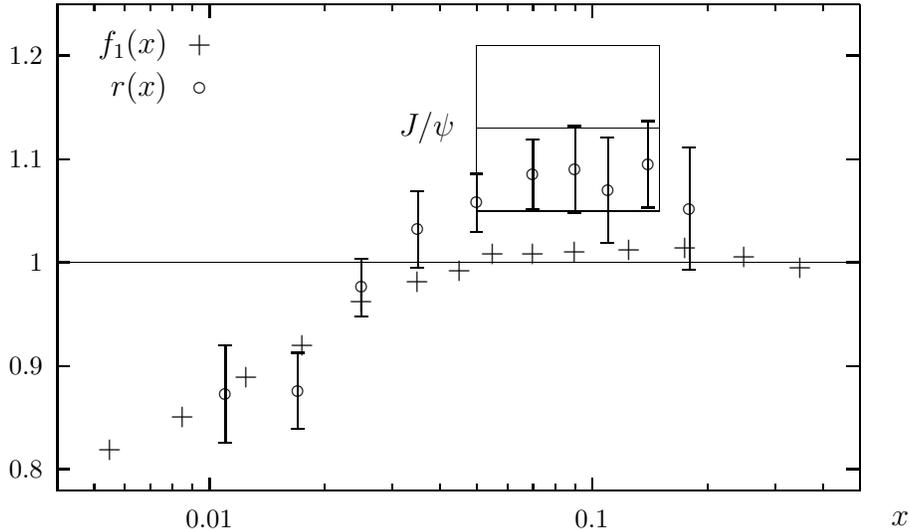

The extraction corresponding to Eq.~(\ref{ratio_e}) is given in Table~\ref{ratio_t} 
for each $x$ together with the corresponding $\langle Q^2\rangle$. 
We plot the ratio $r(x)$ in Fig.~\ref{ratio_f} together with $f_1(x)$. One 
observes a signal for a stronger antishadowing in the gluon case ($\approx$8\%) 
than in the quark case ($\approx$1\%). In the shadowing region, there is no evidence 
for a difference between quarks and gluons. It is, however, presently impossible to 
draw a firm conclusion in this region.

It turns out that the uncertainty in the right hand side of Eq.~(\ref{ratio_e}) comes 
predominantly from the error in $f_2$ so that, in Table~\ref{ratio_t}, we report 
the uncertainty
$$
\delta r={\Delta f_2\over f_3-{\alpha_S\over 2\pi}{\Delta^C\over F_2^C}},
$$
where $\Delta f_2$ is the experimental statistical error on $f_2$. The systematic 
error is in the process of being assessed. 

There are also theoretical approximations. Neglect of NLL corrections together 
with the use of Eq.~(\ref{approx}) lead to an approximation presumably better 
than 20\% for an absolute extraction of gluon density. Because the ratio $r(x)$ 
remains close to 1 in the whole range of $x$ being considered, they are hopefully 
even far better for the ratio we are evaluating.
\medskip

The ratio $r$ has also been extracted from $J/\psi$ electroproduction~\cite{nmc92} 
in the range $0.05\le x\le 0.15$, with the averaged result 
$\langle r\rangle=1.13\pm 0.08$. In Fig.~\ref{ratio_f}, we show this value as a 
rectangle the sides of which are the $x$ range of the data and the errorbar in 
$\langle r\rangle$. One can see that the two different extractions are in good 
agreement. 
\medskip

Using the result on $r$, one can estimate the consequence on 
$\Delta R=R^{\rm Sn}-R^{\rm C}$, the difference of the ratios of longitudinal 
to transverse virtual photon cross sections. In principle measurement of 
$\sigma_L$ at small $x$ provides a direct information on $G(x)$. This 
measurement is, however, difficult, therefore experimental analysis of $\sigma_L$ 
is still going on in the small $x$ range. Here we examine the problem by going in 
the opposite direction, that is to see whether the extracted ratio of gluon 
distributions, $r(x)$, is compatible with the NMC measurement of 
$\Delta R$~\cite{nmc}.

From the expression of $R$,
\begin{equation}
R={F_L+{4 M_N^2x^2\over Q^2}F_2\over F_2-F_L}\approx{F_L\over F_2-F_L},
\end{equation}
assuming $\Delta R\ll R$ (see below), one finds
\begin{equation}
\Delta R\approx R(1+R)\left({1\over f_1}{F_L^{\rm Sn}\over F_L^{C}}-1\right).
\end{equation}
At next to leading order, $F_L$ is given by~\cite{roberts}
\begin{equation}
F_L(x)={\alpha_S\over 2\pi}x^2\int_x^A{dy\over y^3}
\left({8\over3} F_2(y)+2\sum e_i^2(1-x/y)\,y G(y)\right).
\end{equation}
At small $x$, the integrals can be approximated in a way similar to the method 
leading to Eq.~(\ref{approx}). One has~\cite{roberts}
\begin{equation}
F_L(x)\approx {\alpha_S\over 2\pi}\left({4\over3} F_2(2x)
+{2\sum e_i^2\over 5.85}\,2.5x\,G(2.5x)\right).
\end{equation}
The $F_2$-terms are smaller than the $G$-terms, but they are not negligible in 
the whole $x$ range. A detailed analysis would require a better knowledge 
of absolute values of $F_2$'s and $G$'s, so we restrict ourselves to an estimate 
for $\Delta R$ by dropping the $F_2$-terms to simply end with
\begin{equation}
\left.{\Delta R\over R(1+R)}\right|_{x,Q^2}\approx {r(2.5x,Q^2)\over f_1(x,Q^2)}-1.
\end{equation}
Simple parametrizations of $r$ and $f_1$ shows that this quantity varies very 
weakly with $x$ in the range $0.01\le x\le 0.1$. It is about 0.08$\pm$0.01. $R$ 
is about 0.1{--}0.3 for the nucleon case in this $x$-range (and for the $Q^2$ 
of the data), so that one verifies the above statement that $\Delta R\ll R$, and 
one gets $0.01\le\Delta R\le 0.03$. This is compatible with the new data from 
NMC~\cite{nmc}. We note, however, that the present precision of the data on 
$\Delta R$ unfortunatly does not give much constraint on the extracted ratio 
$r(x)$.  
\bigskip

The present analysis can be compared with previous analysis~\cite{gup92,kha95} 
of $J/\psi$-production in hadronic collisions~\cite{ald91}. In an older 
attempt~\cite{gup92}, the experimental data have been all accounted for 
by gluon shadowing. The resulting ratio,
$G^{\rm Sn}/G^{\rm C}\approx1+(0.021\pm 0.001)\ln x_2\ln 118/12$,
would be much below our determined ratio in Fig.~\ref{ratio_f} for $x\ge 0.01$. 
Recently, the same data have been reanalysed~\cite{kha95} on the basis of a 
Fock state decomposition of the charmonium states with an essential contribution 
from the $|(c\bar{c})_8 g\rangle$ component, supposed to give the interaction of 
the precharmonium state soon after formation. This new analysis can explain the 
FNAL data without shadowing of the gluon distribution at variance with our analysis.
\medskip

The shape of $r(x)$ obtained in this paper has been anticipated in~\cite{fra90} on 
the basis of the fulfilment of the momentum sum-rule. In order to get a momentum 
fraction for the gluons $\approx$2\% higher in the Calcium nucleus than in a 
free nucleon, shadowing in the gluon distribution at $x<0.05$  was predicted to 
be compensated by a noticeable antishadowing around $x=0.1$. We can infer that 
the general feature discussed in~\cite{fra90} are in qualitative agreement with 
the data. The sum-rule fraction for the shadowing$+$antishadowing region, 
$x<x_0\approx 0.25$, is
\begin{equation}
\gamma_G={\int_0^{0.25}dx\,(r(x)-1)\,xG(x)\over\int_0^1 dx\,xG(x)}\approx2\%,
\end{equation}
in tin to carbon. This value is close to the estimated enhancement of the gluon 
fraction in Ca to D. Let us note for completeness that the contribution from 
$x>0.25$ to the momentum fraction carried by gluons is not totally negligible, 
in spite of the small gluon density, $G(x)$, at large $x$. Typically, a mean 
depletion of 10\% in the interval $0.25\le x\le0.5$ (EMC effect for gluons) 
diminishes $\gamma_G$ from 2\% to 1\%. 
\medskip

The range of $x$ values accessible via the NMC $F_2$-measurements 
$0.01\le x\le 0.2$ is sufficient to estimate the importance of shadowing 
for the minijet production rate at RHIC. One needs, however, much smaller 
$x$ information, down to $x\approx 10^{-4}$, for heavy ion collision at 
LHC. To cover this range, HERA with nuclear beams would be very helpful. 
The scaling violation analysis shown here may have difficulties to be used 
since the standard evolution of the $F_2$ structure function, even with 
NLL corrections included, may be modified by non-leading twist terms 
coming from the merging of partons\footnote{Let us mention that 
this non-linear effect rather than being a limitation to gluon density 
analysis is to be seen as one of the very challenging effects to be studied at 
HERA.}~\cite{mue86}. It is therefore important to extend the inclusive 
measurements of $F_2$ to jet, open charm and charmonium productions.
\bigskip

We gratefully acknowledge two members of NMC, Antje Br\H ull and Andreas 
M\H ucklich, for providing us the data and related informations. We have also 
benefitted from discussions with participants of the working group on nuclei 
at HERA.

\end{document}